\documentstyle[12pt]{article}   

\large
\newcommand{\tiN}{\raisebox{-6.5pt}{$\displaystyle
\stackrel{\displaystyle N}{\sim}$}}
\newcommand{\tiM}{\raisebox{-6.5pt}{$\displaystyle
\stackrel{\displaystyle M}{\sim}$}}
\newcommand{\beq}{\begin{equation}}
\newcommand{\eeq}{\end{equation}}

\makeatletter
\@addtoreset{equation}{section}
\makeatother

\title{Generalized boundary conditions for general relativity for the
asymptotically flat case in terms of Ashtekar's variables}
\author{T. Thiemann\thanks{hiemann@phys.psu.edu} \\
       Institute for Theoretical Physics, RWTH Aachen,\\
       D-52074 Aachen, Germany}
\date{{\small Preprint PITHA 93-31, August 93}}

\begin{document}

\maketitle                     

\begin{abstract}

There is a gap that has been left open since the formulation of general
relativity in terms of Ashtekar's new variables namely the
treatment of asymptotically flat field configurations that are general
enough to be able to define the generators of the Lorentz subgroup of the
asymptotical Poincar\'e group. While such a formulation already exists for the
old geometrodynamical variables, up to now only the generators
of the translation subgroup could be defined because the function spaces
of the fields considered earlier are taken too special. The transcription
of the framework from the ADM variables to Ashtekar's variables turns out not
to be
straightforward due to the freedom to choose the internal SO(3) frame at
spatial infinity and due to the fact that the non-trivial reality conditions
of the Ashtekar framework reenter the game when imposing suitable boundary
conditions on the fields and the Lagrange multipliers.
\end{abstract}

\section{Introduction}

Since the advent of the new canonical variables introduced by Ashtekar
(\cite{1}) the majority of related contributions have dealt, within the
canonical treatment of gravity, with the case of a compact topology
of the initial data hypersurface without boundary because it is
technically simpler and although the major problem of general relativity,
to define what {\em is} and how to {\em construct} an observable, is even
more severe than for the case of an asymptotically flat topology since
one then does not even have access to the well-known surface observables
at spatial infinity (spi, not to be confused with the universal structure
at spatial infinity which is called Spi (\cite{3}).\\
Recently, the Killing-reduced model of spherically symmetric gravity
with or without a Maxwell-field was shown to be a system for which the
programme of canonical quantization can be carried out completely
(\cite{4}). Since this is an isolated sytem which is of physical interest
only in the asymptotically flat context, the author was forced to consider
the most general (spherically symmetric) field configurations and thus
was lead out of the limitations given by the fall-off behaviour of the
fields as proposed in (\cite{1}). He found a genuinely complex
ADM-momentum as defined for Ashtekar's formulation of general relativity
even after the reality conditions have been imposed and figured out
that this can be traced back to the fact that the symmetry generator
corresponding to the vector constraint as defined in
\cite{1} is not differentiable if one considers fall-off conditions
appropriate for the case of spherical symmetry. The problem was cured for
spherical symmetry and by the way the more general fall-off conditions, valid
for the full theory, were found, too.\\
This paper reports about these results :\\
\\
In section 2 we review the framework given in \cite{2} since there we find
an explicit and consistent definition of the generators of the asymptotic
Poincar\'e group. The calculations and definitions that follow will all be
based on and motivated by the results of that paper.

In section 3 we propose a transcription of the definition of the function
spaces of the basic fields from the geometrodynamical to Ashtekar's
variables. The challenge is that there is now more asymptotic structure
available since we are a priori free to choose the internal SO(3) frame.
It turns out, however, that in order to recover the ADM-energy as given
in \cite{2} requires to fix the asymptotic internal frame. This is
satisfactory because an 'SO(3)-charge' should not play any role in general
relativity.\\
The correction for the generating functional for the symplectomorphism
from the old to the new variables compared to that in \cite{1} is
obtained.

In section 4 we derive the generators of the Poincar\'e group for the new
variables.
These are constructed from the constraint functionals by the requirement
that they should be finite and functionally differentiable (together
with the symplectic form) on the full phase space.
Now, the virtue of the new variables is that the constraint functions turn
out
to be {\em polynomial}. Furthermore, with the function spaces considered in
\cite{1}, the translation generators turn out to be also polynomial in the
new variables. The main result is that if one wants to include boosts and
rotations
into the framework, {\em the Poincar\'e generators necessarily become
nonpolynomial}. The constraint functions themselves, of course, remain
polynomial so that the main advantage of the new variables is not
invalidated.\\
However, this nonpolynomial nature of the Poincar\'e generators has an
important consequence : it is not possible anymore, as stated frequently
(\cite{1}), to consider degenrate metrics (actually there is already a problem
with degenerate metrics when imposing the reality conditions for Ashtekar's
variables which involve the nonpolynomial spin-connection).\\
Altogether, what we do is just to do the symplectomorphism from the ADM
phase space to the Ashtekar phase space correctly which in particular
implies that on the constraint surface defined by the Gauss constraint
all the results of $\cite{2}$ remain valid.

Section 5 checks that the Poincar\'e generators defined satisfy the correct
gauge algebra, that they are Dirac observables (\cite{5}) and that on the
constraint surface one recovers the Lie algebra of the Poincare group
modulo the well known supertranslation ambiguity.

Throughout it is assumed that the reader is familiar with the formulation
of general relativity in terms of the new variables. For a review the
reader is referred to \cite{1}. We are using the abstract index notation.

\section{Geometrodynamics in the asymptotically flat context}

We give here mainly a compact account of the results obtained in \cite{2}
for later reference and since we want to compare the old and new formulation
of general relativity. For more details the reader is referred to that
paper :
\\
\\
The basic variables of the canonical formulation of general relativity for
geometrodynamics are the intrinsic metric (or 1st fundamental form) $q_{ab}$
on the initial data hypersurface $\Sigma$ and its extrinsic curvature
(or 2nd fundamental form) $K_{ab}$ (more precisely the second basic
variable is the momentum conjugate to $q_{ab}$, $p^{ab}:=(\det(q))^{1/2}
(q^{ac} q^{bd}-q^{ab} q^{cd})K_{cd},\;\mbox{where}\;q^{ab}$ is the inverse
of $q_{ab}$). As usual for the discussion of asymptotic flatness (\cite{6}),
we will assume the topology of $\Sigma$ to be such that it is homeomorphic
to the union of a compact set with a collection of 'ends' (i.e. asymptotic
regions) where each end is homeomorphic to the complement of a compact ball
in $R^3$. In the sequel we restrict ourselves to one end and mean by
$\partial\Sigma\;\equiv\;S^2$ only spatial infinity of that end. The
treatment of the remaining regions is identical.\\
Let us introduce a local frame in the neighbourhood of spi, denoted by
$\{\bar{x}^a\}$, and we abbreviate the asymptotic 'spherical' variables by
$r^2:=\bar{x}^a\bar{x}^b\delta_{ab},\phi^a:=\bar{x}^a/r\;\mbox{where}\;
\phi^a$ coordinatizes spi (topologically $S^2$). We do not take this frame to
be necessarily cartesian, but we assume that the transition matrix
$(\partial (x_{cart})^a/\partial \bar{x}^b$ mediating to the cartesian one,
$(x_{cart})^a$, is of order $O(1)$.\\
Motivated by the appearence of the Schwarzschild
solution in these coordinates one defines
\beq q_{ab}\rightarrow \bar{q}_{ab}+\frac{f_{ab}}{r}+O(1/r^2) \eeq
as $r\rightarrow\infty$ and it is assumed that the tensors $\bar{q}_{ab},
f_{ab}$ are smooth on spi ($\bar{q}_{ab}$ is the euclidian metric in
the asymptotic frame and it is assumed to be nondynamic, i.e. it is taken, as
well as $\partial\Sigma$ as the '1st order part of the asymptotic structure
at spi' (\cite{3})).\\
In order to derive the asymptotic behaviour of $p^{ab}$ one seeks to make
the symplectic structure
\beq \Omega:=\int_\Sigma d^3x\; dp^{ab}(x)\wedge dq_{ab}(x) \eeq
well-defined (recall that for an integral curve $\gamma$ with parameter t
on the phase space the gradient of $q_{ab}$ is defined by
$d q_{ab}(x)[(\partial/\partial t)_{|\gamma}]:=\dot{q}_{ab}(x)_{|\gamma}$
and similar
for the momentum variable). Furthermore, one wants to have the ADM-momentum
to be non-vanishing (see formula (2.9)). This implies that the leading
order part of $p^{ab}$ must be $O(1/r^2)$. However, then the symplectic
structure is logarithmically divergent in general. This can be cured by
imposing that the leading order parts of the dynamical parts of the
basic variables should have opposite parity. Since for an asymptotic
translation lapse and shift are even functions at spi of order 1, the
only way to have the ADM-4-momentum non-vanishing is to impose that
$f_{ab}$ is even while $h^{ab}$ is odd (and smooth) at spi where
\beq p^{ab}\rightarrow \frac{h^{ab}}{r^2}+O(1/r^3) \; \eeq
(in order to see this, recall that the surface integrals are {\em first}
evaluated
at finite r and {\em then} one carries out the limit $r\to\infty$. We have
then $dS_a:=\frac{1}{2}\epsilon_{abc}d\bar{x}^b\wedge d\bar{x}^c=n_a r^2 d\mu\;
\mbox{where}\;
\mu$ is the standard measure at spi and $n_a$ is the outward unit normal
at spi, i.e. $dS_a$ is 'odd').\\
The constraint functionals of source-free general relativity in the old
ADM-variables are given by
\begin{eqnarray}
V_a[N^a] & = & \int_\Sigma d^3x N^a(-2 D_b p^b_a) \; ,\\
C[N] & = & \int_\Sigma d^3x N(\frac{1}{\sqrt{\det(q)}}(p^{ab}p_{ab}
-\frac{1}{2}(p^a_a)^2)-\sqrt{\det(q)}(^{(3)}R))
\end{eqnarray}
called, respectively, the vector and the scalar constraint. Here we have
introduced the components tangential ($N^a$, the shift vector) and
perpendicular (N, the lapse function) to the hypersurface $\Sigma$ of the
foliation-defining vector field (with parameter t) and $^{(3)}R$ is the
scalar curvature of $(\Sigma,q_{ab})$. It is understood that all indices
are raised and lowered with the intrinsic metric whose unique torsion-free
covariant differential is given by D.\\
The vanishing of these constraint
functionals defines the contraint surface $\bar{\Gamma}$ of the phase
space. Since these constraints turn out to be first class in Dirac's
terminology (more precisely, $\bar{\Gamma}$ is a coisotropic submanifold
of $\Gamma$) they are to generate gauge transformations and so
define the reduced phase space $\hat{\Gamma}$ by identifying points in
$\bar{\Gamma}$ which lie on the same flow line of the Hamiltonian vector
fields associated to the constraint functionals. It follows that one needs
to compute Poisson-brackets with these constraint-functionals.\\
In order that one can compute Poisson-brackets of these constraint
functionals with various functions on the phase space, they have to be\\
1) finite i.e. the integrals have to converge, \\
2) functionally differentiable.\\
Inserting the fall-off behaviour
of our 2 basic fields into the equations (2.4), (2.5) one discovers that the
integrals diverge, in general, if one not restricts the fall-off of the
'Lagrange-multipliers' $(N^a,N)$, too. There is no problem if one chooses
them of, say, compact support but then the ADM-4 momentum (to be defined
shortly) vanishes identically. In order to account for the generators of
the asymptotic Poincar\'e group, the most general behaviour of lapse and
shift that one would like to incorporate into the analysis is as follows :
\begin{eqnarray}
N^a &\rightarrow& a^a+\bar{\eta}^a\;_{bc}\varphi^b \bar{x}^c
+\mbox{supertranslations} \nonumber\\
N &\rightarrow& a+\bar{q}_{ab}\beta^a \bar{x}^b+\mbox{supertranslations} \; .
\end{eqnarray}
Here $(a^a,a)$ are the parameters of the translation subgroup, $\varphi^a$
are rotation angles and $\beta^a$ are boost angles. These are vectors in the
asymptotic $R^3$, i.e. constants, while the supertranslation parameters
(to be specified in more detail shortly) are genuinely angle-dependent
functions on $S^2$ (for more information about the supertranslation
ambiguity, see ref. \cite{3},\cite{7}).\\
However, now the above generators of gauge transformations fail to be
(manifestly) well-defined, nor are they functionally differentiable. This
can be cured by using the following procedure : \\
Vary the basic fields and obtain expressions proportional to $\delta q_{ab},
\delta p^{ab}$. In doing this one picks up a surface integral. If the
volume term of the variation is well-defined, it gives the searched for
functional derivative of the functional we are seeking to make well defined.
Subtract the surface
term from the variation of the original constraint. If then finally the
obtained surface term turns out to be exact, i.e. can be written as the
variation of an ordinary surface integral, one has obtained this expression
which is functionally differentiable and, if one is lucky, is already
(manifestly) finite.\\
Let us exemplify this for the vector constraint : $D_b p^b_a\;\mbox{is}\;
O(1/r^3)$ and even, so even for an asymptotic translation the integral in
(2.4) diverges logarithmically. Upon variation we obtain
\beq \delta V_a[N^a]=\int_\Sigma d^3x ({\cal L}_{\vec{N}}q_{ab}\delta p^{ab}
-{\cal L}_{\vec{N}}p^{ab}\delta q_{ab})+2\int_{\partial\Sigma}dS_b N^a \delta
p^b_a \; . \eeq
The volume part is obviously already finite : since $\vec{N}$ is an
asymptotic Killing vector of $\bar{q}_{ab}$, we have
${\cal L}_{\vec{N}}q_{ab}=O(1/r^2)$ odd and $O(1/r)$ even for an asymptotic
translation or rotation respectively while $\delta p^{ab}=O(1/r^2)$ odd.
Hence
${\cal L}_{\vec{N}}p^{ab}=O(1/r^3)$ even and $O(1/r^2)$ odd for an asymptotic
translation or rotation respectively while $\delta q_{ab}=O(1/r)$ even.
Hence, the integrand is either $O(1/r^4)$ even or $O(1/r^3)$ odd and so
converges. \\
On the other hand $N^a\delta p^b_a=\delta(N^a p^b_a)$, so the surface
term is indeed exact. Hence our candidate for a finite and functionally
differentiable vector constraint arises from subtracting the corresponding
counterterm :
\beq H_a[N^a]:=V_a[N^a]+2\int_{\partial\Sigma}dS_b N^a p^b_a
=\int_\Sigma d^3x {\cal L}_{\vec{N}}q_{ab} p^{ab} \eeq
where we have carried out an integration by parts in the second step. This
functional is differentiable by construction and luckily already finite
since (by the same argument as above) the integrand is either $O(1/r^4)$
even or $O(1/r^3)$ odd. The surface term in (2.8) defines the ADM-momentum
for an asymptotically constant shift, the integrand is $O(1)$ even and
hence does not vanish :
\beq 2\int_{\partial\Sigma}dS_b N^a p^b_a=:a^a P_a \eeq
wile for an asymptotic rotation we obtain
\beq 2\int_{\partial\Sigma}dS_b N^a p^b_a=:\varphi^a\bar{\eta}_{abc}\bar{x}^b
P^c \eeq
which qualifies $J^a:=\bar{\eta}_{abc}\bar{x}^b P^c$ as the asymptotic angular
momentum.\\
For the scalar constraint one proceeds similarily. The difference is that
due to the appearence of second spatial derivatives in $^{(3)} R$
one has to do an integration by parts twice in order to
arrive at a well-defined variation of the scalar constraint. The final
result is given by (\cite{2})
\beq H[N]:=C[N]+2\int_{\partial\Sigma} dS_d\sqrt{\det(q)}q^{ac} q^{bd}
[N \bar{D}_{[c} q_{b]a}-(q_{a[b}-\bar{q}_{a[b})\bar{D}_{c]} N] \eeq
where $\bar{D}$ is the unique torsion-free covariant differential compatible
with $\bar{q}_{ab}$. The first part of the surface integral is called
the ADM-energy for an asymptotic time-translation
\beq E:=2\int_{\partial\Sigma} dS_d\sqrt{\det(q)}q^{ac} q^{bd}
\bar{D}_{[c} q_{b]a} \eeq
while the second part then vanishes. Otherwise we obtain the boost-generator
\beq \beta_e K^e:=\beta^e(E \bar{x}^e
-2\int_{\partial\Sigma} dS_d\sqrt{\det(q)}q^{ac} q^{bd}
(q_{a[b}-\bar{q}_{a[b})\delta_{c]}^e) \; .\eeq
Note that since gauge transformations are those for which the generators
$H,H_a$ induce identity transformations at spi on the constraint surface,
they have to vanish then. Thus, only the odd part of the supertranslation
parameters are non-vanishing in this case
(the integrand without the Lagrange multipliers is even).
In the seqel we will therefore consider the following fall-off behaviour
of the Lagrange-multipliers :
\begin{eqnarray}
N^a:=T^a+R^a+S^a=T^a+\bar{\eta}^a\;_{bc}\varphi^b\bar{x}^c+S^a, \nonumber\\
N:=T+B+S=T+\beta_a\bar{x}^a+S
\end{eqnarray}
where $(T,T^a)$ are O(1) even, i.e. generate an asymptotic translation in
4-dimensional space, $B,R^a$ are $O(r)$ odd, i.e. generate asymptotic
boosts and rotations and finally $(S,S^a)$ are $O(1)$ and odd at least in
leading order and thus generate the odd supertranslations. Moreover, the
vector field $(T+B)\bar{n}^a+(T^a+R^a)$ is required to be a (10-parameter)
Killing vector field of the asymptotic spacetime metric
$\bar{g}_{ab}=\bar{q}_{ab}-\bar{n}_a \bar{n}_b$ which implies, in particular,
that ${\cal L}_{\vec{T}}q_{ab}$ is $O(1/r^2)$ odd and that
${\cal L}_{\vec{R}}q_{ab}\;\mbox{and}\;{\cal L}_{\vec{S}}q_{ab}$ are $O(1/r)$
even. These properties will be used in the following.

\section{Connection dynamics in the asymptotically flat context}

We begin by transcribing the boundary conditions imposed on the
ADM-variables to the new variables. In the following, latin letters from the
beginning of the alphabet will denote tensor indices while latin letters from
the middle of the alphabet will denote (internal) SO(3) indices.\\
Recalling that the triad 1-form
is the square-root of the 3-metric, we expect the following fall-off
behaviour
\beq e_a^i\rightarrow\bar{e}_a^i+\frac{f_a^i(\phi^b)}{r}+O(1/r^2) \eeq
where we have called the triad of the asymptotic 3-metric at spi
$\bar{e}_a^i$.
In order to investigate what the relation between the smooth tensors
$f_{ab}\;\mbox{and}\;f_a^i$ on spi is, we compute the 3-metric
\beq q_{ab}=\delta_{ij}e_a^i e_b^j=\bar{q}_{ab}
+2\frac{\delta_{ij}\bar{e}_{(a}^i f_{b)}^j}{r}+O(1/r^2) \eeq
from which we infer
\beq f_{ab}=2\delta_{ij}\bar{e}_{(a}^i f_{b)}^j \; .\eeq
This has the solution
\beq f_a^i=(\frac{1}{2}f_{ab}+\bar{\eta}_{abc}v^c)\bar{e}^b_i \eeq
where $v^c$ is a smooth vector field at spi.\\
Since the parity of $f_{ab}$ is even, $\bar{e}_a^i\;\mbox{and}\;f_a^i$ must
have equal parity. Note that we do not require $\bar{e}_a^i$ to be a constant
(even if the asymptotic frame is cartesian).
Also, the vector field $v^a$ is not constrained at all
up to now.\\
Since (if the Gauss-constraint is satisfied) $K_a^i=K_{ab} e^b_i$ we need
the asymptotic formula for the inverse triad. From the requirement
$q^{ac}q_{cb}=\bar{q}^a_b$ we obtain unambiguosly
\beq q^{ab}=\bar{q}^{ab}-\bar{q}^{ac}\bar{q}^{bd}\frac{f_{cd}}{r}+O(1/r^2)
\eeq whence
\beq e^a_i=q^{ab}e_b^i=\bar{e}^a_i-\frac{\frac{1}{2}\bar{q}^{ac}f_{bc}
-\bar{\eta}^a\;_{bc}v^c}{r}\bar{e}^b_i \; .\eeq
Accordingly we obtain
\beq K_a^i=\frac{h_{ab}(\phi^c)\bar{e}^b_i}{r^2}+O(1/r^3) \eeq
and the smooth tensor $h_{ab}$ at spi has odd parity.\\
Finally, in order to derive the asymptotic formula for the
Ashtekar-connection we need to determine the asymptotics for the
spin-connection $\omega_a\;^i_j$. First of all we extend the covariant
derivative D to generalized tensors in the usual way, e.g.
\beq D_a e_b^i:=\bar{D}_a e_b^i-\Gamma^c\;_{ab}e_c^i+\omega_a\;^i_j e_b^j=0
\eeq
where $\bar{D}$ is the covariant differential, extended to generalized
tensors, that annihilates $\bar{e}_a^i$ and $\bar{q}_{ab}$. Note that this
implies that the spin- and metric-connections of $e_a^i$ are $O(1/r^2)$ while
those of $\bar{e}_a^i$ are only $O(1/r)$.\\
{}From the torsion-freeness requirement
$\bar{D}\wedge e^i+\omega^i_j\wedge e^j=0$ we easily infer ($\bar{D}$ acts on
tensor - {\em and} SO(3) indices !)
\beq \frac{r^2\bar{D}_{[a}\frac{f_{b]}^i}{r}+\lambda_{[a}^i\;_j
\bar{e}_{b]}^j}{r^2}+O(1/r^3)=0 \eeq
where we have denoted the leading order part of the spin-connection by
$\lambda_a\;^i_j$. Since $\bar{D}_b f_a^i\;\mbox{and}\;\bar{e}_a^i$ have
opposite
parity, this formula can only hold if $\lambda_a\;^i_j$ has odd parity.\\
Note :\\
This may also be checked explicitely : inverting formula (3.8), the analytic
expression for the spin-connection in terms of the triads is given by
(we define, as usual, $\Gamma_a^i:=1/2\epsilon^{ijk}\omega_a\;_{jk}$)
\beq \Gamma_a^i=-\frac{1}{2}\epsilon^{ijk}e^b_j(2\bar{D}_{[a}e_{b]}^k
+e^c_k e_a^l\bar{D}_c e_b^l) \eeq
from which we immediately derive
\beq \lambda_a^i=-r^2\frac{1}{2}\epsilon^{ijk}\bar{e}^b_j(2\bar{D}_{[a}
\frac{f_{b]}^k}{r}
+\bar{e}^c_k\bar{e}_a^l\bar{D}_c \frac{f_b^l}{r}) \;. \eeq
Since this is an even polynomial in $\bar{e}_a^i\;\mbox{and}\;f_a^i$
(both of which have equal parity), homogenous of 1st degree in the spatial
derivatives, and r is an even function at spi, expression (3.11) is
altogether
an odd quantity. Hence, it follows directly from the formula for the
Ashtekar-connection, $A_a^i=\Gamma_a^i+i K_a^i$, that its leading order
part $G_a^i$ where
\beq A_a^i=\frac{G_a^i}{r^2}+O(1/r^3) \eeq
has definite parity if and only if the leading order part of $K_a^i$
is odd, i.e. if  and only if $\bar{e}_a^i$ is even.\\
Since the 2nd variable of the new canonical formulation of general relativity
is given by $E^a_i:=\sqrt{\det(q)}e^a_i$, i.e.
\beq E^a_i=\bar{E}^a_i-\sqrt{\det(\bar{q})}\frac{1/2\bar{q}^{ac}f_{bc}
-\bar{\eta}^a\;_{bc}v^c+\bar{q}^{bc}f_{bc}\bar{q}^a_b}{r}\bar{e}^b_i+O(1/r^2)
=: \bar{E}^a_i+\frac{F_a^i}{r}+O(1/r^2) \; .\eeq
we see that the next to leading
order part $F_a^i$ has definite parity if and only if we set $v^a=0$. This
parity is, fortunately, even if we want $G_a^i$ to have definite (odd)
parity. Hence, it is possible to choose parity conditions in such a way
that $A_a^i E^a_i$ is asymptotically an odd scalar density without violating
that $E^a_i\;\mbox{and}\;A_a^i$ 'have to come from $q_{ab}\;\mbox{and}\;
K_{ab}$', however, these
parity conditions are rather awkward : they imply that $E^a_i$ becomes
asymptotically a {\em pseudo}-vector density while $A_a^i$ is asymptotically
a {\em true} covector. Moreover, parity conditions cannot be reversed
if one wants to make the Ashtekar formulation meaningful in the
asymptotically flat context, just as for the old ADM-formulation.\\
\\
As one can check, the symplectic form is already finite thanks to our
parity conditions (we set the gravitational coupling constant equal to
one in the sequel) :
\beq \Omega:=\int_\Sigma d^3x (-i) dE^a_i\wedge dA_a^i
\rightarrow \int_\Sigma d^3x
[\frac{dF^a_i\wedge dG_a^i}{r^3}+O(1/r^4)] \; .\eeq
Recalling that the Ashtekar-action arises from a {\em complex} canonical
transformation, we ask if it is still true that its generating functional
is given by $\int_\Sigma d^3x E^a_i\Gamma_a^i$ which was the case for
the boundary conditions discussed by Ashtekar (\cite{1}). But it is
obvious that this functional diverges logarithmically without further
specification. Indeed, using the explicit
expression (3.10) we compute
\beq 2E^a_i\Gamma_a^i=\epsilon^{abc}e_c^i\bar{D}_b e_a^i
\rightarrow\frac{r^2\epsilon^{abc}
\bar{e}_c^i\bar{D}_b\frac{f_a^i}{r}}{r^2}+O(1/r^3) \eeq
and the first term vanishes since $\bar{D}_a\bar{e}_b^i=0\mbox{ and }f_{ab}$
is symmetric, while the second has no definite parity. We can cure this
simply by applying the following trick : using that $\bar{D}_a\bar{e}_b^i=0$
we can write
\begin{eqnarray}
 & & 2 E^a_i\Gamma_a^i=\epsilon^{abc}e_c^i \bar{D}_b e_a^i \nonumber\\
 & = & \epsilon^{abc}e_c^i \bar{D}_b(e_a^i-\bar{e}_a^i) \nonumber\\
 & = & \bar{D}_b(\epsilon^{abc}e_c^i (e_a^i-\bar{e}_a^i))
-\epsilon^{abc}(\bar{D}_b e_c^i)(e_a^i-\bar{e}_a^i) \nonumber\\
 & = & \partial_b(\epsilon^{abc}e_c^i (e_a^i-\bar{e}_a^i))
 -\epsilon^{abc}(\bar{D}_b e_c^i)(e_a^i-\bar{e}_a^i) \; .
\end{eqnarray}
The 2nd term in the last line is manifestly $O(1/r^3)$ and odd, so gives
a convergent integral, while the first yields a surface integral.
Accordingly, we propose to define the generating functional for the
spin connection by
\beq  \int_\Sigma d^3x E^a_i\Gamma_a^i-\frac{1}{2}
\int_{\partial\Sigma} dS_b(\epsilon^{abc}e_c^i (e_a^i-\bar{e}_a^i)) \eeq
and we can quickly check that this expression is indeed functionally
differentiable and that the functional derivative is precisely the
spin-connection. To see this, vary the integrand of the volume integral
to obtain 2 terms
\beq \delta(E^a_i\Gamma_a^i)=\Gamma_a^i\delta E^a_i+E^a_i\delta\Gamma_a^i \;.
\eeq
The first is already the required one while the second can be written
(after tedious calculations) as
\beq E^a_i\delta\Gamma_a^i
=\frac{1}{2}\partial_a(\epsilon^{abc}e_b^i\delta e_c^i) \;. \eeq
Varying the surface term on the other hand (recall that the asymptotic triad
is non-dynamical, i.e. $\delta\bar{e}_a^i=0$) yields
\begin{eqnarray}
& &  -\frac{1}{2} \int_{\partial\Sigma} dS_b(\epsilon^{abc}
[(e_a^i-\bar{e}_a^i)\delta e_c^i+e_c^i \delta e_a^i] \nonumber\\
& = & -\frac{1}{2} \int_{\partial\Sigma} dS_a \epsilon^{abc}
e_b^i \delta e_c^i
\end{eqnarray}
where the first term in the first line of (3.20) has dropped out because it
is finite and odd. Comparing this result with eq. (3.19) we observe that the
2 terms exactly cancel.

\section{The Poincar\'e-generators for the new variables}

The final task is now to make the constraints convergent and functionally
differentiable. We begin with the Gauss-constraint :
\beq {\cal G}_i:=\bar{D}_a E^a_i+\epsilon_{ijk}A_a^j E^a_k \; .\eeq
The first term is $O(1/r^2)$ while the second is $O(1/r^3)$ and the whole
expression is odd. Hence we need only worry about the first term whose
variation, when integrated against the Lagrange-multiplier $\Lambda^i$,
is just given by
\beq \int_{\partial\Sigma}dS_a\Lambda^i\delta E^a_i \;.\eeq
Recalling that $\Lambda^i$ is nothing else than the time component of the
self-dual part of the 4-dimensional spin-connection, we require that it
is $O(1/r^2)$ for a symmetry transformation. Hence, the
Gauss-constraint is already finite and functionally differentiable if we
further require that $\Lambda^i$ be even. This parity condition is also
consistent with the interpretation of $\Lambda^i=T^a\;^{(4)}A_a$ because
it is the contraction of an odd vector with an odd covector. There is
thus no need for the SO(3)-charge
\beq Q:=\int_\Sigma dS_a E^a_i\Lambda^i \eeq
which then vanishes anyway due to parity and whose variation is also zero.
Note, however, that for model systems which do not have access to
parity (e.g. \cite{4}) the SO(3)-charge is non-vanishing and gives a
spurious observable which is fortunately unnecessary (the SO(3)-charge
should play no role in general relativity).\\
The treatment of the vector and scalar constraint turns out to be much
more difficult and differs from the one given in \cite{1}. This is to be
expected because the generating functional (3.17) of the canonical
transformation
to the new canonical variables as given here is different from the
one given in \cite{1}. It gives rise to additional terms in the action,
both in the volume and the surface part.\\
It turns out to be the easiest strategy in finding the necessary
modifications to take the old ADM-action in its manifestly finite and
functionally differentiable version and then to express it in terms of
the new variables. We cannot expect it to remain well-defined because there
are now 'more' variables. However, using the gauge constraint, we must be
able to restrict the variations of the constraints to those of the
ADM phase space. This is the outline of the idea, let us now come to the
technical part of it.\\
Let us start with the expression (2.8), the generator of asymptotical
spatial translations and rotations. We have
\begin{eqnarray}
H_a[N^a] & = & \int_{\Sigma}d^3x p^{ab}{\cal L}_{\vec{N}}q_{ab}
           =   -\int_{\Sigma}d^3x p_{ab}{\cal L}_{\vec{N}}q^{ab}\nonumber\\
& = & -\int_{\Sigma}d^3x p_{ab}{\cal L}_{\vec{N}}(\frac{E^a_i E^b_i}{\det(q)})
\nonumber\\
& = & -\int_{\Sigma}d^3x \frac{p_{ab}}{\det(q)}[2E^b_i{\cal L}_{\vec{N}}E^a_i
-E^a_i E^b_i(E^{-1})_c^j{\cal L}_{\vec{N}}E^c_j] \nonumber\\
& = & -2\int_\Sigma d^3x [(K_{ab}-K q_{ab})e^b_i+K e_a^i]
{\cal L}_{\vec{N}}E^a_i \nonumber\\
& = & -2\int_\Sigma d^3x K_a^i{\cal L}_{\vec{N}}E^a_i
\end{eqnarray}
where we have used the elementary fact that $p^{ab}{\cal L}_{\vec{N}}q_{ab}
=p^{ab}{\cal L}_{\vec{N}}(q_{ac}q_{bd}q^{cd})=2p^{ab}{\cal L}_{\vec{N}}q_{ab}
+p_{ab}{\cal L}_{\vec{N}}q^{ab}=-p_{ab}{\cal L}_{\vec{N}}q^{ab}$ that
$\det(q)=\det(E^a_i)$ and the definition of the momentum conjugate to the
3-metric in terms of the extrinsic curvature
(it seems as if up to now we have only used differential geometric identities
but actually we made use of the Gauss-constraint in order to replace the
variable
$K_{ab}\;\mbox{by}\;K_a^i$).\\
\\
Finally we use the reality conditions in order
to write (3.4) in terms of the Ashtekar-connection :
\beq H_a[N^a]=2i\int_\Sigma d^3x (A_a^i-\Gamma_a^i){\cal L}_{\vec{N}}E^a_i
\; . \eeq
Noting that $E^a_i$ is a covector density of weight one, we have
${\cal L}_{\vec{N}}E^a_i=\bar{D}_b N^b E^a_i+N^b\bar{D}_b E^a_i-E^a_i
\bar{D}_b N^a$ and this expression is obviously $O(1/r)$ odd for an
asymptotic translation while it is O(1) even for an asymptotic rotation
(note that this expression is a genuine generalization to arbitrary
asymptotic frames of the formula for the cartesian frame since $\bar{D}$ acts
also on internal indices; for the cartesian frame the metric- and spin
connection vanish asymptotically so that we recover the usual formula in
this case; also, for usual tensors there is no difference. The motivation
for this generalization is that we want to keep $\partial_a E^b_i=O(1/r^2)$
when passing to an arbitrary asymptotic frame).
Hence the integral (4.5) diverges in either of these cases ! To see how
this comes about, we compute the Lie-derivative of the twice densitized
inverse asymptotic metric in terms of the electric fields
\beq 0={\cal L}_{\vec{N}}(\det(\bar{q})\bar{q}^{ab})=
2\bar{E}^{(a}_i {\cal L}_{\vec{N}}\bar{E}^{b)}_i \eeq
which vanishes since $\vec{N}$ is an asymptotic Killing vector of
$\bar{q}_{ab}$. Thus, we see the source of the divergence in (4.5) : the
fact that $\vec{N}$ is an asymptotic Killing-vector {\em does not} imply
that $\bar{E}^a_i$ is also Lie-annihilated but only that it is
asymptotically rotated in the tangent space :
\beq {\cal L}_{\vec{N}}\bar{E}^a_i=\bar{\eta}^a\;_{bc}\xi^b \bar{E}^c_i\;
\mbox{i.e.}\; \xi_a=\frac{1}{2\det(\bar{q})}\bar{\eta}_{abc}\bar{E}^b_i
\bar{E}^c_i \eeq
is $O(1/r)$ odd or O(1) even for an asymptotic translation or rotation
respectively. In order to isolate the divergence in (4.5) we write
\begin{eqnarray}
& & K_a^i {\cal L}_{\vec{N}}E^a_i=K_a^i {\cal L}_{\vec{N}}(E^a_i-\bar{E}^a_i)
+K_a^i {\cal L}_{\vec{N}}\bar{E}^a_i \nonumber\\
& = & K_a^i\bar{\eta}^a\;_{bc}\xi^b\bar{E}^c_i+\mbox{finite} \nonumber\\
& = & K_a^i\epsilon^{ijk}\bar{E}^a_j\bar{E}^b_k\frac{\xi_b}
{\sqrt{\det(\bar{q})}}+\mbox{finite} \nonumber\\
& = & ([\epsilon^{ijk}K_a^j\bar{E}^a_k]\bar{E}^b_i+K_a^i\epsilon^{ijk}
[\bar{E}^a_j-E^a_j]\bar{E}^b_k)
\frac{\xi_b}{\sqrt{\det(\bar{q})}}+\mbox{finite} \nonumber\\
& = & \frac{1}{i}{\cal G}_i\bar{E}^b_i\frac{\xi_b}{\sqrt{\det(\bar{q})}}
+\mbox{finite}
\end{eqnarray}
where we absorbed terms that are either $O(1/r^4)$ even or $O(1/r^3)$ odd
into the expression 'finite'. Thus we managed to peel out the contribution
that causes trouble in (4.5) : as expected, it is something proportional to
the Gauss-constraint and hence plays no role on the constraint-surface.\\
It is then motivated to consider as a candidate for a well-defined
symmetry-generator corresponding to the vector constraint (we use the
same label)
\beq H_a[N^a]:=2i\int_\Sigma d^3x[(A_a^i-\Gamma_a^i){\cal L}_{\vec{N}}E^a_i
-\frac{1}{\sqrt{\det(q)}}E^a_i\xi_a{\cal G}_i] \eeq
where $\xi_a:=1/(2\det(q))\eta_{abc}E^b_i{\cal L}_{\vec{N}}E^c_i$. The reader
might worry about the highly non-polynomial appearence of (4.9) but, as
we will show, {\em the constraints} remain polynomial and this is the
important thing to keep when thinking of quantizing general relativity
(\cite{1}).\\
Note that we have 'unbarred' everything in (4.9) as compared to (4.8) which
we justify now by direct computation, thereby showing that (4.9) is
well-defined and differentiable. By simply using the definitions we
verify that the integrand of (4.9) is given by
\beq (E^{-1})^i_b(A_a^i-\Gamma_a^i){\cal L}_{\vec{N}}(\det(q)q^{ab}) \eeq
which is $O(1/r^3)$ odd or $O(1/r^4)$ even repectively and thus yields a
convergent integral thanks to the fact that $\vec{N}$ is an asymptotic
Killing vector (recall the discussion at the end of the previous section).
Thus, we have already established finiteness.\\
In order to show that (4.9) is also already functionally differentiable,
we first display the relation of (4.9) with the constraints. We have
\begin{eqnarray}
-K_a^i{\cal L}_{\vec{N}}E^a_i & = & -{\cal L}_{\vec{N}}(K_a^i E^a_i)
+E^a_i{\cal L}_{\vec{N}}K_a^i \nonumber\\
& = &(N^b \bar{D}_b K_a^i+K_b^i\bar{D}_a N^b)E^a_i-\bar{D}_b(N^b K_a^i E^a_i)
\nonumber\\
& = & N^a(E^b_i\bar{D}_a K_b^i-\bar{D}_b(K_a^i E^b_i))
-2\bar{D}_{[b}(N^b K_{a]}^i E^a_i) \nonumber\\
& = & -iN^a(2 E^b_i\bar{D}_{[a} A_{b]}^i-A_a^i\bar{D}_b E^b_i)
-iN^a(2 E^b_i\bar{D}_{[a} \Gamma_{b]}^i-\Gamma_a^i\bar{D}_b E^b_i)
\nonumber\\
& & +2i\bar{D}_{[b}(N^b [A_{a]}^i-\Gamma_{a]}]E^a_i) \; .
\end{eqnarray}
Due to the definition of the spin-connection, we have $0=D_a E^a_i=\bar{D}_a
E^a_i+\epsilon_{ijk}\Gamma_a^j E^a_k$ so that the bracket of the 2nd term in
the last equation can be written
\beq 2 E^b_i\bar{D}_{[a} \Gamma_{b]}^i-\Gamma_a^i\bar{D}_b E^b_i
=E^b_i(2\bar{D}_{[a} \Gamma_{b]}^i+\epsilon_{ijk}\Gamma_a^j\Gamma_b^k)
=E^b_i R_{ab}^i\equiv 0 \eeq
due to the first Bianchi-identity. Accordingly, (4.11) can be rewritten in
the form
\beq
-iN^a(F_{ab}^i E^b_i-A_a^i{\cal G}_i)
+2i\bar{D}_{[b}(N^b [A_{a]}^i-\Gamma_{a]}]E^a_i) \eeq
where $1/2 F$ is the curvature of the Ashtekar-connection and we have used
the definition of the Gauss-constraint (4.1). Since the variation of the
action with respect to $\Lambda^i$ yields the Gauss-constraint, the variation
with respect to the shift-vector field results in the
diffeomorphism-contraint, i.e. the 1st bracket in (4.13), for a shift
corresponding to the generator of a gauge transformation (i.e. the shift is
at most a supertranslation at spi). Hence the polynomial form of the
constraints is not spoiled by the nonpolynomial term in the full vector
contraint
\beq H_a[N^a]=-2i\int_\Sigma d^3x[N^a(F_{ab}^i E^b_i-A_a^i{\cal G}_i)
+\frac{1}{\sqrt{\det(q)}}E^a_i\xi_a{\cal G}_i]
-4i\int_{\partial\Sigma}dS_a[A_b^i-\Gamma_b^i]N^{[b}E^{a]}_i) \; .\eeq
Note that the surface term {\em is not} Ashtekar's expression (\cite{1})
which is just the one given in (4.14) but {\em without} the second term
including the spin-connection. It has the advantage of
giving (modulo reality conditions) manifestly the ADM-momentum.
For a pure translation the surface term in (4.14) reduces to
Ashtekar's expression but, as we will show, this is not true for our
boundary conditions when including the asymptotic Lorentz group.\\
Let us now check functional differentiability of (4.14). Looking at the
surface term that is picked up when varying (4.14) we obtain
\begin{eqnarray}
& & \delta H_a[N^a]_{|\partial\Sigma} \nonumber\\
 & = & \int_{\partial\Sigma}
[-4 i dS_{[a}\delta A_{b]}^i N^a E^b_i+2i dS_b A_a^i\delta E^b_i N^a
\nonumber\\
& & -2i(\det(q))^{-1/2}(E^a_i\xi^a)\delta E^b_i dS_b
-i(\det(q))^{-3/2}{\cal G}_i E^a_i \nonumber\\
& & \eta_{abc}E^b_j{\cal L}_{\vec{N}}
\delta E^c_j  4idS_{[a}(\delta A_{b]}^i-\delta\Gamma_{b]}^i)E^b_i
\nonumber\\
& & -4i dS_a(A_b^i-\Gamma_b^i) N^{[b}\delta E^{a]}] \; .
\end{eqnarray}
The first line comes from the variation of the polynomial part of the
volume term, the second from its nonpolynomial part and the last line
arises from varying the surface term. We observe that the first term of the
1st line cancels against the first term in the first bracket of the last
line. Furthermore, the 2nd term of the first line as well as the complete
last bracket of the last line and the term proportional to the
Gauss-constraint in the middle line vanish identically irrespective of the
nature of the tranformation due to parity or fall-off. Recalling the
definition of $\xi^a$ we thus end up with
\beq \delta H_a[N^a]=-2i\int_{\partial\Sigma}dS_a[\frac{1}{2[\det(q)]^{3/2}}
E^b_i\eta_{bcd}E^c_j{\cal L}_{\vec{N}}E^d_j-2N^{[b}\delta(E^{a]}_i\Gamma_b^i)]
\; .\eeq
By using the explicit expression (3.10) for the spin-connection in terms of
the triads, it is easy to show that
\begin{eqnarray}
E^a_i\Gamma_a^i & = & \frac{1}{2}\epsilon^{abc}e_c^i\bar{D}_b e_a^i,
\nonumber\\
E^a_i\Gamma_b^i & = & \frac{1}{2}\epsilon^{acd}(\bar{D}_c q_{bd}
+e_c^i\bar{D}_b e_d^i) \; .
\end{eqnarray}
Since the terms in which the underivated parts are varied vanish due to
parity or fall-off, we arrive after simple algebra at
\begin{eqnarray}
\delta H_a[N^a] & = & i\int_{\partial\Sigma}(\det(q))^{1/2} dS_a
[\eta^{acd}(\frac{1}{2}D_c N_d q^{ef}\delta q_{ef} \nonumber\\
 & & -N^b(D_d\delta q_{bc}-e_c^i D_b\delta e_d))
+\eta^{bcd}(e_b^i\delta e^a_i D_c N_d-N^a e_d^i D_c\delta e_b^i]
\end{eqnarray}
where we have used $E^a_i=\sqrt{q}^{1/2}e^a_i$ in order to write
everything in terms of variation of the triads and we could replace
$\bar{D}\delta f\;\mbox{by}\; D\delta f$ for any f because the additional
contributions included in D fall-off one order faster while (4.18) is
anyway either $O(1)$ even or $O(r)$ odd. \\
We will now discuss the 3 possibilities for the fall-off of the shift :\\
Case 1) supertranslation :\\
(4.18) is $O(1)$ and odd, so vanishes identically.\\
Case 2) translation :\\
We have now $D_a N_b=O(1/r^2)$ so there are no contributions from these
terms in (4.18), hence
\begin{eqnarray}
\delta H_a[N^a] & = & -i\int_{\partial\Sigma}(\det(q))^{1/2} dS_a
[\eta^{acd}{ef}N^b(D_d\delta q_{bc}-e_c^i D_b\delta e_d)
+\eta^{bcd}N^a e_d^i D_c\delta e_b^i] \nonumber\\
& = & i\int_{\partial\Sigma}(\det(q))^{1/2} dS_a [(\eta^{bcd} N^a-\eta^{acd}
N^b)e_d^i D_b\delta e_c^i-\eta^{acd}N^b D_d\delta q_{bc}]\; .
\end{eqnarray}
The last term can be written
\beq (\det(q))^{1/2}dS_a\eta^{acd}N^b D\delta q_{bc}=dx^c\wedge dx^d
D_d(N^b\delta q_{bc})+O(1/r)=d\wedge(N^b\delta q_{bc} dx^c)+O(1/r) \eeq
and thus drops out since $\partial\partial\Sigma=\emptyset$. The first term
on the other hand can be cast into the form
\begin{eqnarray}
& &(\det(q))^{1/2}dS_a (\eta^{bcd} N^a-\eta^{acd}N^b)e_d^i D_b\delta e_c^i
\nonumber\\
& =& (\det(q))^{1/2}dS_a D_b[(\eta^{bcd} N^a
-\eta^{acd}N^b)e_d^i]\delta e_c^i \nonumber\\
 & & +O(1/r)=(\det(q))^{1/2}dS_a D_b\eta^{abc}f_c+O(1/r)=d\wedge f+O(1/r)
\end{eqnarray}
where we have defined the 1-form
$f_e:=-1/2\eta_{eab}(\eta^{bcd} N^a-\eta^{acd}N^b)e_d^i\delta e_c^i$. Hence,
(4.19) is just the integral of an exact 1-form over an exact chain and thus
vanishes identically.\\
case 3) rotation :\\
We now exploit $D_a N_b=\eta_{abc}\xi^c,\; D_a\xi^c=O(1/r^2)$. The idea is
to integrate by parts in (4.19) and to order terms proportional to $N^a$ or
$D_a N^b$. We thus have, using $D_a e_b^i=0$,
\begin{eqnarray}
\delta H_a[N^a] & = & i\int_{\partial\Sigma}(\det(q))^{1/2} dS_a
[\eta^{acd}(\frac{1}{2}q^{ef}\delta q_{ef}D_c N_d+\delta q_{bc}D_d N^b
\nonumber\\
& & -e_c^i \delta e_d D_b N^b)
+\eta^{bcd}(e_b^i\delta e^a_i D_c N_d+e_d^i\delta e_b^i D_c N^a)\nonumber\\
& & +\eta^{acd}(-D_d(\delta q_{bc}N^b)+D_b(e_c^i \delta e_d N^b)
-\eta^{bcd} D_c(e_d^i\delta e_b^i D_c N^a)] \;.
\end{eqnarray}
We can write the last two brackets as folows :
\begin{eqnarray}
& & (\det(q))^{1/2} dS_a[-\eta^{acd}D_d(\delta q_{bc}N^b)
-D_b([\eta^{acd} N^b-\eta^{bcd}N^a]e_d^i \delta e_c)] \nonumber\\
& = & (\det(q))^{1/2} dS_a[-\eta^{acd}D_d(\delta q_{bc}N^b)
-\eta^{abe}D_b(\eta_{efg}\eta^{fcd}N^g e_d^i \delta e_c)] \nonumber\\
& = & (\det(q))^{1/2} dS_a\eta^{abe}D_b[\delta q_{ec}N^c
-\eta_{efg}\eta^{fcd}N^g e_d^i \delta e_c]
\end{eqnarray}
and thus displays it as an exact 2-form. Since we have regularized the
surface integrals in such a way (see the remark in section 2) that they are
to be evaluated at finite r and then one takes the limit, we see that due
to the fact that the sphere at finite r is also without boundary, the
integral over (4.23) drops out of (4.22). We are thus left with the first
two brackets in (4.22). These turn out to cancel each other algebraically
when using the above expression for $D_a N_b$ :
\begin{eqnarray}
& & \eta^{acd}(\frac{1}{2}q^{ef}\delta q_{ef}D_c N_d+\delta q_{bc}D_d N^b
-e_c^i \delta e_d D_b N^b)
+\eta^{bcd}(e_b^i\delta e^a_i D_c N_d+e_d^i\delta e_b^i D_c N^a)\nonumber\\
& = & \xi^a q^{ef}\delta q_{ef}+2q^{[a}_e q^{c]}_f \xi^f q^{eb}\delta q_{bc}
-2\xi^b e^a_i \delta e_b^i
+2q^{[d}_e q^{b]}_f\xi^f q^{ae} e_d^i\delta e_b^i \nonumber\\
& = & q^{ab} \xi^c\delta q_{bc}-\xi^b e^a_i \delta e_b^i
-\xi^d q^{ab} e_d^i\delta e_b^i \nonumber\\
& = & q^{ab}\xi^c(\delta q_{bc}-e_c^i\delta e_b^i)-\xi^b e^a_i \delta e_b^i
\nonumber\\
& = & q^{ab}\xi^c e_b^i\delta e_c^i-\xi^b e^a_i \delta e_b^i\equiv 0
\end{eqnarray}
where in the last step we used $\delta q_{ab}=2e_{(a}^i\delta e_{b)}^i$.\\
Thus we proved that (4.14) is finite, functionally differentiable,
reduces to the ADM-momentum on
the constraint surface and its (weakly) vanishing part is polynomial in the
basic variables. It therefore satisfies all requirements that one wishes to
impose on a symmetry generator corresponding to momentum. We will prove
in the next section that (4.14) generates the correct symmetries and gauge
transformations.\\
\\
Note : \\In principle one could have found the correct version (4.14) of the
vector-constraint in terms of Ashtekar's variables also simply by carefully
implementing the Gauss-constraint thus displaying (4.14) {\em exactly} as the
old ADM-momentum expression after doing the complex canonical transformation
which leads to Ashtekar's theory : starting again from (2.8) we have
\begin{eqnarray}
\int_\Sigma d^3x p^{ab}{\cal L}_{\vec{N}} q_{ab}
& = & -\int_\Sigma d^3x p_{ab}{\cal L}_{\vec{N}} q^{ab} \nonumber\\
& = & -\int_\Sigma d^3x\sqrt{\det(q)}(K_{ab}-K q_{ab})
{\cal L}_{\vec{N}} q^{ab} \nonumber\\
& = & -\int_\Sigma d^3x\frac{1}{\sqrt{\det(q)}}(K_{ab}\det(q)
{\cal L}_{\vec{N}} q^{ab}-K{\cal L}_{\vec{N}}\det(q)) \nonumber\\
& = & -\int_\Sigma d^3x\frac{1}{\sqrt{\det(q)}}K_{(ab)}
{\cal L}_{\vec{N}}(\det(q) q^{ab})
\end{eqnarray}
and it is important to see that only the symmetric part of the tensor
$K_{ab}$ contributes to (4.25). The asymmetric part on the other hand is
directly related to the Gauss constraint : using the definition of $A_a^i,
\; K_a^i:=K_{ab}e^b_i\;\mbox{and}\;D_a E^a_i=0$ we have
\beq {\cal G}_i=i\sqrt{\det(q)}\epsilon_{ijk}e^b_j e^a_k K_{ab}=
-i\epsilon^{abc} K_{ab} e_c^i \eeq
and thus we conclude
\begin{eqnarray}
\int_\Sigma d^3x p^{ab}{\cal L}_{\vec{N}} q_{ab}
& = & -2\int_\Sigma d^3x\frac{1}{\sqrt{\det(q)}}K_{(ab)}
{\cal L}_{\vec{N}}(E^a_i)E^b_i \nonumber\\
& = & -2\int_\Sigma d^3x [K_{ab}-K_{[ab]}]
{\cal L}_{\vec{N}}(E^a_i)e^b_i \nonumber\\
& = & -2\int_\Sigma d^3x [K_a^i-\frac{1}{2}\epsilon_{abc}
\epsilon^{dec} K_{de}e^b_i] {\cal L}_{\vec{N}}(E^a_i) \nonumber\\
& = & -2\int_\Sigma d^3x [K_a^i-\frac{i}{2}
\epsilon^{ijk} E^k_a {\cal G}_j] {\cal L}_{\vec{N}}(E^a_i) \nonumber\\
& = & 2i\int_\Sigma d^3x [K_a^i{\cal L}_{\vec{N}}(E^a_i)+\frac{1}{2}
(\epsilon^{ijk}{\cal L}_{\vec{N}}(E^a_j) E^k_a) {\cal G}_i]
\end{eqnarray}
which is precisely (4.14).\\
\\
\\
We finally come to discuss energy. Again we follow the strategy to take the
old ADM expression (2.11), write it in terms of Ashtekar's variables and look
for the necessary modifications. Writing the Ashtekar-connection in terms
of the spin-connection and the extrinsic curvature,
it is easy to show that (compare \cite{1}) the ADM- scalar constraint
turns out to be
\begin{eqnarray}
C[\tiN] & = & \int_\Sigma d^3x \tiN[F_{ab}^i\epsilon_{ijk}E^a_j E^b_k+2 D_a
(E^a_i{\cal G}_i))] \nonumber\\
& = & \int_\Sigma d^3x \tiN[F_{ab}^i\epsilon_{ijk}E^a_j E^b_k+2 \bar{D}_a
(E^a_i{\cal G}_i))]
\end{eqnarray}
(we used the fact that the divergence of a vector density is independent
of the metric connection)
and in \cite{2} the necessary counterterms are obtained simply by integrating
by parts this expression an appropriate number of times in order to arrive
at a manifestly finite expression. We follow this approach here.\\
We observe that the part proportional to $\bar{D}A$ in the first term
and the 2nd term in (4.28) are either $O(1/r^3)$ even or $O(1/r^2)$ odd so
they diverge while the remainder is already finite. We are now going to
integrate by parts both divergent terms :
\begin{eqnarray}
C[\tiN] & = &\int_\Sigma d^3x[-2A_b^i\bar{D}_a(\tiN\epsilon_{ijk}E^a_j E^b_k)
-2 E^a_i{\cal G}_i D_a\tiN+\tiN A_a^j A_b^k\epsilon_{ijk} E^a_m E^b_n
\epsilon^{imn}] \nonumber\\
& & + \int_{\partial\Sigma} dS_a 2\tiN [A_b^i\epsilon_{ijk}E^a_j E^b_k
+E^a_i{\cal G}^i] \nonumber\\
 & = & \int_\Sigma d^3x[-2 (D_a\tiN) E^a_i\bar{D}_b E^b_i
+\tiN(-2A_b^i\epsilon_{ijk}\bar{D}_a(E^a_j E^b_k)
+ A_a^j A_b^k\epsilon_{ijk} E^a_m E^b_n\epsilon^{imn})] \nonumber\\
& & + \int_{\partial\Sigma} dS_a 2\tiN E^a_i\bar{D}_b E^b
\end{eqnarray}
where we have used the definition of the Gauss-constraint (4.1). While the
term proportional to $\tiN$ in the volume integral is already manifestly
convergent, the term proportional $D_a\tiN$ is not. However, we write
in the same term $\bar{D}_b E^b_i=\bar{D}_b (E^b_i-\bar{E}^b_i)$ and
integrate by parts again. Note that $\bar{D}_b D_a\tiN=(\det(q))^{-1/2}
\bar{D}_a\bar{D}_b N+O(1/r^2)$ is of order $O(1/r^2)$ at least, even for a
boost ($\partial(x_{cart})^b/\partial\bar{x}^a$ is of order one and in the
cartesian frame $(x_{cart})^a$ the assertion is obvious). We thus obtain
\begin{eqnarray}
C[\tiN] & = & \int_\Sigma d^3x[-2\bar{D}_b((D_a\tiN)E^a_i)(E^b_i-\bar{E}^b_i)
+\tiN(-2A_b^i\epsilon_{ijk}\bar{D}_a(E^a_j E^b_k) \nonumber\\
& & + A_a^j A_b^k\epsilon_{ijk} E^a_m E^b_n\epsilon^{imn})]
+ \int_{\partial\Sigma} dS_a 2[\tiN E^a_i\bar{D}_b E^b \nonumber\\
& & -(D_b\tiN) E^b_i(E^a_i-\bar{E}^a_i)]
\end{eqnarray}
and now the volume integral is manifestly convergent and thus gives rise to
the ansatz for the correct symmetry generator :
\beq H[\tiN]:=C[\tiN]-\int_{\partial\Sigma} dS_a 2[\tiN E^a_i\bar{D}_b E^b
-(D_b\tiN) E^b_i(E^a_i-\bar{E}^a_i)] \; .\eeq
In fact, it is easy to prove that (4.31) is already differerentiable.
The boundary contribution of the volume term is given by
\begin{eqnarray}
\delta C[\tiN]_{|boundary\;term} & = & 2\int_{\partial\Sigma} dS_a\tiN
[\epsilon_{ijk}E^a_j E^b_k\delta A_b^i+\delta(E^a_i{\cal G}_i)]
\nonumber\\
& &  -\{2\int_\Sigma d^3x \bar{D}_a\tiN \delta(E^a_i{\cal G}_i)
\}_{|boundary\;term} \\
& = &  2\int_{\partial\Sigma} dS_a\tiN
[\epsilon_{ijk}E^a_j E^b_k\delta A_b^i+E^a_i(\bar{D}_b\delta E^b_i
\nonumber\\
& & +\epsilon_{ijk} E^b_k\delta A_b^j]-\{2
\int_\Sigma d^3x \bar{D}_a\tiN E^a_i\bar{D}_b\delta(E^b_i\}_{|boundary\;term}
\nonumber\\
& = & 2\int_{\partial\Sigma} dS_a
[\tiN E^a_i\bar{D}_b\delta E^b_i-D_b\tiN E^b_i\delta E^a_i]
\end{eqnarray}
where we could neglect all terms that are of order $O(1/r^3)$ (in particular
we could replace $\bar{D}$ by D).
Hence we have for the
variation of the surface integral in (4.31)
\beq -\delta C[\tiN]_{|boundary\;term}-2\int_{\partial\Sigma} dS_a
[\tiN \delta E^a_i\bar{D}_b E^b-(D_b\tiN)\delta E^b_i(E^a_i-\bar{E}^a_i)]
\eeq
and the integrand of the latter integral is easily seen to be of order
$O(1/r^2)$ even or $O(1/r^3)$ odd and thus vanishes identically.\\
This completes the proof that expression (4.31) is differentiable.\\
Let us check that the boundary term in (4.31) is indeed the expression (2.11)
given in section 2. We have
\begin{eqnarray}
E^a_i\bar{D}_b E^b_i & = & \sqrt{\det(q)}e^a_i\bar{D}_b(\sqrt{\det(q)}e^b_i)
\nonumber\\
& = & \det(q)(E^a_i\bar{D}_b e^b_i+\frac{1}{2}q^{ab}q^{cd}\bar{D}_b q_{cd})
\nonumber\\
& = & \det(q)q^{ac}(e_c^i\bar{D}_b e^b_i+\frac{1}{2}q^{bd}\bar{D}_c q_{bd})
\nonumber\\
& = & \det(q)q^{ac}q^{bd}(-e_d^i\bar{D}_b e_c^i+\frac{1}{2}\bar{D}_c q_{bd})
\end{eqnarray}
whence (remember that $\tiN\sqrt{\det(q)}=N$ is the lapse), by using the
the Gauss-constraint,
\begin{eqnarray}
& & -\int_{\partial\Sigma} dS_a 2\tiN A_b^i\epsilon_{ijk}E^a_j E^b_k
\nonumber\\
& \approx & -\int_{\partial\Sigma} dS_a 2\tiN E^a_i\bar{D}_b E^b
\nonumber\\
& = & \int_{\partial\Sigma} dS_d N\sqrt{\det(q)}q^{ac}q^{bd}
(2e_a^i\bar{D}_c e_b^i-\bar{D}_c q_{bd}) \nonumber\\
& = & \int_{\partial\Sigma} dS_d N\sqrt{\det(q)}q^{ac}q^{bd}
([\bar{D}_c q_{ab}-\bar{D}_c q_{bd}]+2e_{[a}^i\bar{D}_{|c|} e_{b]}^i)
\nonumber\\
& = & E+\int_{\partial\Sigma} dS_d \sqrt{\det(q)}q^{ac}q^{bd}
2e_{[a}^i\bar{D}_{|c|} e_{b]}^i
\end{eqnarray}
and the integrand of the second term in (4.36) can be shown to vanish to
2nd order in $1/r$ by virtue of the symmetry of $f_{ab}$ (recall (2.1)) :
\begin{eqnarray}
e_{[a}^i\bar{D}_{|c|} e_{b]}^i &=& (\bar{e}_{[a}^i+\frac{f_{[a}^i}{r}+O(1/r))
(\bar{D}_{|c|} \frac{f_{b]}^i}{r}+O(1/r^3))\nonumber\\
& = &  \bar{e}_{[a}^i(\bar{e}^d_i\bar{D}_{|c|} \frac{f_{b]d}}{2r}
+\frac{f_{b]d}}{2r}\bar{D}_c\bar{e}^d_i)+O(1/r^3)\nonumber\\
& = &  \bar{q}_{[a}^d\bar{D}_{|c|} \frac{f_{b]d}}{2r}+O(1/r^3)\nonumber\\
& = &  \bar{D}_c \frac{f_{[ba]}}{2r}+O(1/r^3)=O(1/r^3)
\end{eqnarray}
since $\bar{D}\bar{q}_{ab}=\bar{D}\bar{e}^a_i=0$. Hence we have already
reproduced the ADM-energy.\\
As far as the second contribution in (4.31) is concerned we first note
that the $O(1)$ contribution is odd and thus integrates to zero.
Hence we may replace any quantity of the integrand by another term which
differs from the original one by a term which is higher by one order of
$1/r$. We thus write first with a glimpse at (2.11)
\begin{eqnarray}
\int_{\partial\Sigma} dS_a 2 (D_b\tiN) E^b_i(E^a_i-\bar{E}^a_i)
=\int_{\partial\Sigma} dS_a 2 (D_b \tiN)  \nonumber\\
e^b_i(e^a_i-\sqrt{\frac{\det(\bar{q})}{\det(q)}}\bar{e}^a_i)
=\int_{\partial\Sigma} dS_a 2 (D_b \tiN) (q^{ab}
-\sqrt{\frac{\det(\bar{q})}{\det(q)}}e^b_i\bar{e}^a_i)
\end{eqnarray}
and we expand the second term in (4.38) according to (3.1) :
\begin{eqnarray}
& & \sqrt{\frac{\det(\bar{q})}{\det(q)}}e^b_i\bar{e}^a_i)
=\sqrt{\frac{1}{\det(\bar{q}^{-1}q)}}(\bar{e}^b_i
-\bar{q}^{bc}\frac{f_{cd}}{2r}\bar{e}^d_i)\bar{e}^a_i)+O(1/r^2)\nonumber\\
& = & (1-\frac{\bar{q}^{ef}f_{ef}}{2r})(\bar{q}^{ab}
-\bar{q}^{bc}\frac{f_{cd}}{2r}\bar{q}^{ad})+O(1/r^2)\nonumber\\
& = & \bar{q}^{ab}-\frac{1}{2r}(q^{ac}q^{bd}+q^{ab}q^{cd})f_{cd}+O(1/r^2) \;.
\end{eqnarray}
Now, using that $f_{cd}=r(q_{cd}-\bar{q}_{cd})+O(1/r)$ we find that (4.38)
can indeed be written
\begin{eqnarray}
& & \int_{\partial\Sigma} dS_a (D_b \tiN) (q^{ac}q^{bd}-q^{ab}q^{cd})
\frac{f_{cd}}{r} \nonumber\\
& & \int_{\partial\Sigma} dS_d (D_c \tiN) (q^{db}q^{ca}-q^{dc}q^{ba})
\frac{f_{ba}}{r} \nonumber\\
& & \int_{\partial\Sigma} dS_d (D_c \tiN) q^{ab}q^{cd}
(q_{a[b}-\bar{q}_{a[b})D_{c]}N
\end{eqnarray}
and we recover exactly (2.11).\\
Hence we succeeded in giving a differentiable and finite expression which
reproduces the surface terms of the old ADM-theory.

\section{The symmetry-algebra}

We finally come to compute the Poisson-structure of the generators (4.1),
(4.14) and (4.31). The Poisson-structure for Lagrange-multipliers $\Lambda^i,
N^a\;\mbox{and}\;\tiN$ corresponding to pure gauge transformations was
already given in \cite{5}.\\
As in \cite{2} it turns out that even for Lagrange multiplicators
corresponding to symmetries the symmetry algebra just equals the gauge
algebra.\\
Since the transition from the ADM-variables to the new variables is a
canonical transformation, we can copy from \cite{2} the algebra restricted
to the vector- and scalar constraint because in section 4 we showed that
our expressions reduce exactly to the ADM expressions according to this
canonical transformation. The only new brackets are those
including a Gauss-constraint. Let us display the complete variation of the
symmetry generators :
\begin{eqnarray}
\delta{\cal G}_i[\Lambda^i] & = & \int_\Sigma d^3x [(-\bar{D}_a\Lambda^i
+\epsilon_{ijk}\Lambda^j A_a^k)\delta E^a_i
 -\epsilon_{ijk}\Lambda^j E^a_k)\delta A_a^i \nonumber \\
\delta H_a[N^a] & = &\int_\Sigma d^3x [({\cal L}_{\vec{N}} A_a^i
+{\cal L}_{\vec{N}}(\epsilon_{ijk} E_a^j {\cal G}_k)-({\cal G}_i)
\epsilon^{ijk} \nonumber\\
& & ({\cal L}_{\vec{N}}E^b_j E_a^k)E_b^i-(-\bar{D}_a\Lambda[\vec{N}]_i
+\epsilon_{ijk}\Lambda[\vec{N}]^j A_a^k))\delta E^a_i \nonumber\\
& & -({\cal L}_{\vec{N}} E^a_i+\epsilon_{ijk}\Lambda[\vec{N}]_j E^a_k)
\delta A_a^i] \nonumber\\
& & \delta H[\tiN]=\int_\Sigma d^3x[2({\cal D}_b(\tiN \epsilon_{ijk} E^a_j
E^b_k)+\epsilon_{ijk}\Lambda[\tiN]_j E^a_k)\delta A_a^i\nonumber\\
& & -2(\tiN \epsilon_{ijk}F_{ab}^j E^b_k+(\bar{D}_a\tiN){\cal G}_i+
(-\bar{D}_a\Lambda[\tiN]_i \nonumber\\
& & +\epsilon_{ijk}\Lambda[\tiN]^j A_a^k))\delta E^a_i]
\;
\end{eqnarray}
where we have defined
\begin{eqnarray}
\Lambda[\vec{N}]^i & := & \frac{1}{2}\epsilon^{ijk}
({\cal L}_{\vec{N}}E^a_j)E^k_a \; ,\nonumber\\
\Lambda[\tiN]^i & := & -2 (D_a\tiN)E^a_i\; .
\end{eqnarray}
Then it may be checked by explicit calculation that
\begin{eqnarray}
\{{\cal G}_i[\Lambda^i],{\cal G}_j[\Xi^j]\} & = & -i{\cal G}_i[\epsilon_{ijk}
\Lambda^j\Xi^k] \nonumber\\
\{{\cal G}_i[\Lambda^i],H_a[N^a]\} & = & 0 \nonumber\\
\{{\cal G}_i[\Lambda^i],H[\tiN]\} & = & 0 \nonumber\\
\{H_a[M^a],H_b[N^b]\} & = & i H_a[({\cal L}_{\vec{M}}\vec{N})^a] \nonumber\\
\{H_a[M^a],H[\tiN]\} & = & -i H[M^a D_a\tiN] \nonumber\\
\{H[\tiM],H[\tiN]\} & = & -i H[E^a_i E^b_i(\tiM D_b\tiN-\tiN D_b\tiM)]
\end{eqnarray}
according to the rule that
\beq \{A_a^i(x),E^b_b(y)\}=i\delta_a^b\delta^i_j\delta(x,y)\;,
\{A_a^i(x),A_b^j(y)\}=\{E^a_i(x),E^b_j(y)\}=0 \eeq
These equations mean the following (see \cite{2}) : Let
\beq \Phi[N]:={\cal G}_i[N^i]+H_a[N^a]+H[\tiN] \eeq
and compute the Poisson bracket $\{\Phi[M],\Phi[N]\}$ according to (5.4).
Then we have the following combinations (the Lagrange-multiplier of the
Gauss-constraint is always pure gauge, i.e. $O(1/r^2)$ even) :\\
1) M and N both pure gauge (odd supertranslations). Then \newline
${\cal L}_{\vec{M}}
\vec{N},M^a D_a\tiN\;\mbox{and}\;E^a_i E^b_i\tiM D_b\tiN$ are again pure
gauge while $M^i N^j$ is again of even parity when $M^i, N^j$ are, so
the gauge algebra closes, we have a first class system.\\
2) M a symmetry, N pure gauge. \\
Case a) : M a translation. Then ${\cal L}_{\vec{M}}
\vec{N},M^a D_a\tiN\;\mbox{and}\;E^a_i E^b_i\tiM D_b\tiN$ (and their
counterparts obtained by interchanging M and N) are $O(1/r)$ even so the
surface integrals vanish identically.\\
Case b) : M a Lorentz rotation. Then ${\cal L}_{\vec{M}}
\vec{N},M^a D_a\tiN\;\mbox{and}\;E^a_i E^b_i\tiM D_b\tiN$ etc. are
$O(1)$ odd and thus the surface integrals also vanish identically.\\
This means that the Poisson generators have weakly vanishing brackets with
the gauge generators, i.e. they are observables in the sense of Dirac.\\
3) M and N both symmetries. Then ${\cal L}_{\vec{M}}
\vec{N},M^a D_a\tiN\;\mbox{and}\;E^a_i E^b_i\tiM D_b\tiN$ etc. display the
Poisson algebra.\\
Accordingly, we managed to reproduce the results of \cite{2} in the new
variables.


\end{document}